\documentclass[10pt]{article}
\usepackage[a4paper, hmargin=1.55cm, vmargin={1.8cm,1.9cm}, columnsep=0.62cm]{geometry}
\usepackage[T1]{fontenc}
\usepackage{lmodern,microtype}
\usepackage[labelfont=bf,labelsep=period]{caption}
\usepackage{amsmath,amssymb,amsthm,mathtools,bm}
\usepackage{graphicx,booktabs,xcolor,hyperref}
\usepackage[backend=bibtex,style=numeric,sorting=none,date=year]{biblatex}
\title{Random entanglement percolation on realistic quantum networks\thanks{\small Published in \emph{Il Nuovo Cimento C} \textbf{49} (2026) 176. \href{https://doi.org/10.1393/ncc/i2026-26176-2}{doi:10.1393/ncc/i2026-26176-2}}}
\author{Alessandro~Romancino\thanks{\texttt{alessandro.romancino@unipa.it}}\\[0.4em]
\small\textit{Dipartimento di Fisica e Chimica ``E. Segrè'', Università degli Studi di Palermo,}\\
\small\textit{Via Archirafi 36, 90123 Palermo, Italy}
}
\date{April 2026; revised July 2026}

\newcommand{\ket}[1]{\left\lvert#1\right\rangle}

\begin{document}

\maketitle

\begin{abstract}
We study random entanglement percolation in heterogeneous quantum networks, where the singlet-conversion probabilities (SCPs) of the edges are drawn from a probability distribution rather than being fixed. After briefly recalling random classical and random quantum entanglement percolation, we focus on polarization-dependent loss (PDL) as a physical source of random edge entanglement in photonic networks. In this setting, polarization imbalance induces a simple map from the PDL magnitude to the edge SCP. We illustrate this map for representative PDL models and discuss the resulting implications for entanglement percolation.
\end{abstract}

\section{Introduction}

The efficient distribution of entanglement across distant nodes is a central challenge in the development of large-scale quantum networks for the future quantum internet \cite{2018wehner,2008kimble}. Because long-distance quantum channels are inevitably affected by loss and decoherence, several strategies have been proposed to overcome these limitations, including entanglement swapping, entanglement distillation, and quantum repeaters \cite{1999bose,1996deutsch}.

A different approach is entanglement percolation, where long-range entanglement is established by exploiting the topology of a network of partially entangled states \cite{2007acin,2013perseguers,2023menga}. In classical entanglement percolation (CEP), each link is converted into a singlet with the same singlet-conversion probability (SCP), mapping the problem to bond percolation \cite{2007acin}. Quantum entanglement percolation (QEP) improves on this picture by preprocessing the network through local quantum operations, such as $q$-swaps, which modify the effective lattice and can reduce the percolation threshold. Although QEP outperforms CEP on several lattices, the search for the optimal protocol remains open.

Most studies of entanglement percolation assume homogeneous edge entanglement. However, realistic quantum networks are intrinsically heterogeneous: optical links have different lengths and losses, and disorder may naturally generate nonuniform entanglement patterns \cite{2018wehner, 2025meng, 2026cirigliano}. In our previous work, we addressed this more realistic setting by introducing random entanglement percolation, where the SCP associated with each edge is sampled from a probability distribution rather than being fixed.

The aim of the present work is to complement random entanglement percolation with a physically motivated example of random edge distributions. We will show that photonic quantum networks characterized by polarization-dependent loss (PDL) provide a natural example of random edge entanglement, deriving the corresponding law for the distribution of the SCPs in this scenario.

\section{Random Entanglement Percolation}
\label{sec:rep}

In our previous work \cite{2026romancino}, we introduced random entanglement percolation by assigning to each edge of the network a singlet-conversion probability (SCP), denoted by $X$, drawn independently from a probability distribution $f_X$ supported on $[0,1]$. Each link will then have a random probability of being converted to a singlet, otherwise the entanglement is lost. Following the standard procedures in entanglement percolation one converts each link independently \cite{2007acin}. If the initial entanglement is high enough, the percolation transition shows us that the system now percolates, guaranteeing a path of singlet connecting two arbitrary nodes. With random probabilities, we demonstrated that random classical entanglement percolation (RCEP) depends only on this average,
\begin{equation}
    \label{eq:RCEP}
    RCEP(f_X) = CEP(\mu),
\end{equation}
so that, at the macroscopic level, RCEP reduces to standard CEP evaluated at $\mu$. This means that we only need that our random initial entanglement is above the percolation threshold \textit{on average}.

\begin{table*}[t!]
    \caption{Examples of SCP distributions for random
    entanglement percolation. The Bernoulli distribution has probability
    $\mu$ of having SCP equal to $1$ and probability $1-\mu$ of being $0$
    (that is, either a singlet or a product state). Haar-random states
    follow the distribution ${\rm Beta}(1,3)$.}
    \label{tab:examples}
    \centering

    \renewcommand{\arraystretch}{1.5}
    \begin{tabular}{@{}lccc@{}}
        \toprule
        Distribution
        & $\mathbb E[X_{\rm min}]$
        & $C=\dfrac{\mu-\mathbb E[X_{\rm min}]}{\sigma}$
        & $\mu-\mathbb E[X_{\rm min}]=\dfrac{\Delta}{2}$ \\
        \midrule
        Uniform
        & $\mu-\dfrac{\sigma}{\sqrt{3}}$
        & $\dfrac{1}{\sqrt{3}}\approx 0.577$
        & $\dfrac{\sigma}{\sqrt{3}}$ \\

        \addlinespace[3pt]
        Gaussian
        & $\mu-\dfrac{\sigma}{\sqrt{\pi}}$
        & $\dfrac{1}{\sqrt{\pi}}\approx 0.564$
        & $\dfrac{\sigma}{\sqrt{\pi}}$ \\

        \addlinespace[3pt]
        Bernoulli
        & $\mu^2$
        & ---
        & $\mu(1-\mu)=\sigma^2$ \\

        \addlinespace[3pt]
        Symmetric bimodal
        & $\mu-\dfrac{\sigma}{2}$
        & $\dfrac{1}{2}=0.5$
        & $\dfrac{\sigma}{2}$ \\

        \addlinespace[3pt]
        ${\rm Beta}(\alpha,\beta)$
        & $\mu-\dfrac{2B(2\alpha,2\beta)}
          {(\alpha+\beta)B(\alpha,\beta)^2}$
        & $\dfrac{2\sqrt{\alpha+\beta+1}\,B(2\alpha,2\beta)}
          {B(\alpha,\beta)^2\sqrt{\alpha\beta}}$
        & --- \\

        \addlinespace[3pt]
        Haar-random states
        & $\dfrac{1}{7}\approx 0.143$
        & $\dfrac{\sqrt{15}}{7}\approx 0.553$
        & $\dfrac{3}{28}\approx 0.107$ \\
        \bottomrule
    \end{tabular}
\end{table*}

The situation changes for random quantum entanglement percolation (RQEP), where $q$-swap preprocessing is applied before the RCEP step. If $X_1$ and $X_2$ are two independent copies of the edge SCP, the bond generated by a random $q$-swap is $X_{\rm min} := \min(X_1, X_2)$ \cite{1999bose}. Hence, its average is
\begin{equation}
    \label{eq:exp-min}
    \mathbb E[X_{\rm min}] =
    \int_{0}^{1} [1 - F(p)]^2 \,dp = \mu - \frac{1}{2} \Delta,
\end{equation}
where $\Delta = \mathbb E[|X_1 - X_2|]$ is the mean absolute difference of the original distribution and $F$ is the cumulative distribution function associated with the distribution of $X$. Because of Eq.~\eqref{eq:exp-min}, the RQEP protocol now depends also on the shape of the distribution, rather than only on the mean (and the topology of the network). For example, we considered location-scale distributions and found that Eq.~\eqref{eq:exp-min} takes the simple form $\mathbb E[X_{\rm min}] = \mu - C\sigma$, where $\sigma$ is the standard deviation, and the constant $C$ depends only on the shape of the standard distribution. Some examples of distributions are listed in Table~\ref{tab:examples}, where the last column represents the quantum ``penalty'' of RQEP relative to RCEP due to Eq.~\eqref{eq:exp-min}.

\section{Physically Motivated Edge Distributions}

In photonic quantum networks, qubits can be encoded in orthogonal polarization modes \cite{2020wengerowsky}. A natural physical source of randomness in this setting is polarization-dependent loss (PDL), namely the fact that the polarization modes may experience different losses while propagating through a channel \cite{2024kucera}.

Starting from a pure Bell state $\ket{\Phi^+} = (\ket{HH} + \ket{VV}) / \sqrt2$, PDL acts effectively as a local filter on one qubit of a polarization-entangled pair \cite{2018jones,2019kirby}. If the transmitted intensities are $\eta_1 \ge \eta_2$, then, after post-selection, the state is no longer maximally entangled, and the singlet-conversion probability (SCP) is $X = 2\eta_2 / (\eta_1+\eta_2)$. Expressing the imbalance through the PDL magnitude in decibels, $P = 10\log_{10}(\eta_1/\eta_2)$, one obtains
\begin{equation}
    X(P)=\frac{2}{1+10^{P/10}}.
    \label{eq:pdl-scp-map}
\end{equation}
Therefore, any physically motivated distribution of the channel imbalance $P$ induces a corresponding distribution of edge SCPs. If $P$ has density $f_P$, then, using the inverse $P(x)=10\log_{10}[(2-x)/x]$, the induced SCP density is
\begin{equation}
    f_X(x) = f_P(P(x)) \left\lvert\frac{dP}{dx}\right\rvert
    = f_P \left[ 10 \log_{10} \left( \frac{2-x}{x} \right) \right]
    \frac{20}{\ln 10}\frac{1}{x(2-x)}.
    \label{eq:pdl-scp}
\end{equation}

\begin{figure*}[t]
    \centering
    \includegraphics[width=\linewidth]{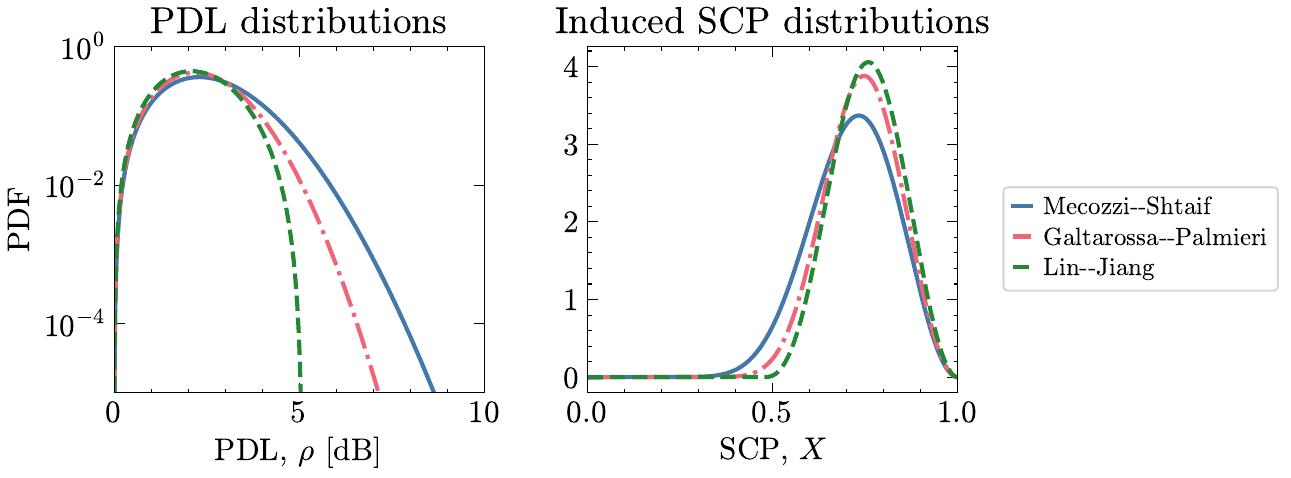}
    \caption{\textit{Left:} probability density of the polarization-dependent loss $P$. \textit{Right:} corresponding induced SCP density obtained from Eq.~\eqref{eq:pdl-scp}. The Mecozzi--Shtaif and Galtarossa--Palmieri curves were matched to the same mean PDL, $\langle P\rangle \approx 2.35\,\mathrm{dB}$, while the Lin--Jiang model used the five-element setting $(0.8,1.2,1.4,1.0,0.7)\,\mathrm{dB}$.}
    \label{fig:pdl-scp-models}
\end{figure*}

Some examples of real-world PDL distributions, shown in Fig.~\ref{fig:pdl-scp-models}, are the Maxwellian model of Mecozzi and Shtaif~\cite{2002mecozzi}, describing the asymptotic regime of many weak independent PDL contributions; the exact accumulated statistics of Galtarossa and Palmieri~\cite{2003galtarossa}; and the finite-element terrestrial-link model of Lin and Jiang~\cite{2024lin}. The corresponding average SCPs, which determine the RCEP threshold, are $\mu_{\rm MS}\approx0.714$, $\mu_{\rm GP}\approx0.739$, and $\mu_{\rm LJ}\approx0.755$. In the weak-PDL regime \cite{2024kucera}, Eq.~\eqref{eq:pdl-scp-map} can be expanded and then averaged to obtain
\begin{equation}
    \mathbb E[X] = 1 - \frac{\ln 10}{20}\,\mathbb E[P] + O(P^3).
\end{equation}

\section{Conclusions}

Random entanglement percolation studied in \cite{2026romancino} is a natural framework for achieving entanglement distribution in heterogeneous networks. The RCEP threshold depends only on the average SCP, while RQEP is sensitive to the whole distribution due to the $q$-swap protocol. Here we have shown that, in photonic quantum networks, polarization-dependent loss provides a natural source of random edge entanglement: the channel imbalance induces a direct map from PDL to SCP. As a consequence, any PDL distribution yields a corresponding distribution of SCPs. We obtain an approximation to the RCEP threshold in the weak-PDL regime.

\section*{Acknowledgments}
The author acknowledges G. Massimo Palma for his assistance during the preparation of this manuscript, as well as Alessandro Candeloro and Sujan Vijayaraj.

\printbibliography

\end{document}